\documentclass[12pt]{article}

\usepackage{epsfig}

{\end{list}}
\newcounter{enumct}

\newcommand{\captive}[1]{\rule{5mm}{0mm}%
\begin{minipage}{150mm}\caption[small]{#1}\end{minipage}}
 
\begin{document}
 
\sloppy

\pagestyle{empty}

\vspace{0.25cm}

\begin{center}
{\LARGE\bf The Vertex Tracker}\\[4mm]
{\LARGE\bf at the $e^+e^-$ Linear Collider}\\[6mm]
{\Large Conceptual Design, Detector R\&D and Physics Performances}\\[4mm]
{\Large for the Next Generation of Silicon Vertex Detectors}\\[10mm]
{\Large Marco Battaglia}\\[3mm]
{\it Department of Physics, High Energy Physics Division}\\[1mm]
{\it University of Helsinki, Finland}\\[1mm]
{\it E-mail: Marco.Battaglia@cern.ch}\\[4mm]
{\Large Massimo Caccia}\\[3mm]
{\it Dipartimento di Fisica}\\[1mm]
{\it Universita' dell'Insubria, Como, Italy }\\[1mm]
{\it E-mail: Massimo.Caccia@mi.infn.it}\\[20mm]
{\bf Abstract}\\[1mm]
\begin{minipage}[t]{140mm}
The $e^+e^-$ linear collider physics programme sets highly demanding 
requirements on the accurate determination of charged particle trajectories 
close to their production point. 
A new generation of Vertex Trackers, based on different technologies of high 
resolution silicon sensors, is being developed to provide the needed 
performances. These
developments are based on the experience with the LEP/SLC vertex detectors 
and on the results of the R\&D programs for the LHC trackers and also define
a further program of R\&D specific to the linear collider applications.
In this paper the present status of the conceptual tracker design, silicon 
detector R\&D and physics studies is discussed.
\end{minipage}\\[5mm]

\rule{160mm}{0.4mm}

\end{center}

\section{Introduction}

With the end of the SLC and LEP operations and the commissioning of the
$B$-factories, the next frontier in $e^+e^-$ collisions is set by a high 
luminosity, high energy linear collider capable of delivering beams 
at centre-of-mass energies in the range 0.3~TeV $< \sqrt{s} <$ 1.0~TeV with 
luminosities in excess to $10^{34}$~cm$^{-2}$~s$^{-1}$.
Expected to be commissioned by the end of the first decade of the new
millennium, the linear collider will complement the physics reach of the 
Tevatron and LHC hadron colliders in the study of the mechanism of 
electro-weak symmetry breaking and in the search for new physics beyond the 
Standard Model~\cite{zerwas}. 
Both precision measurements and particle searches set
stringent requirements on the efficiency and purity of the flavour 
identification of hadronic jets since final states including short-lived $b$ 
and $c$-quarks and $\tau$ leptons are expected to be the main signatures. High
accuracy in the reconstruction of the charged particle trajectories close to 
their production point must be provided by the tracking detectors, in 
particular by the Vertex Tracker located closest to the interaction point, in 
order to perform the reconstruction of the topology of secondary vertices in 
the decay chain of short-lived heavy flavour particles~\cite{heuer}.

If a Higgs boson exists with mass below 150~GeV/c$^2$, as indicated by the 
fit to the present electro-weak data~\cite{Gross}, it will be essential to 
carry out precision measurements of its couplings to different fermion species
as a proof of the mass generation mechanism and to identify its Standard Model
or Supersymmetric nature~\cite{higgs1}. This can be achieved by accurate 
determinations of its decay rate to $b\bar{b}$, $c\bar{c}$, 
$\tau^{+}\tau^{-}$, $W^{+}W^{-}$ and gluon pairs to detect possible deviations
from the Standard Model predictions \cite{higgs2,higgs3}. 
Since the rates for the 
Higgs decay modes into lighter fermions $h^0 \rightarrow c \bar c$, 
$\tau^+ \tau^-$ or into gluon pairs are expected to be only about 10\% or less 
of that for the dominant $h^0 \rightarrow b \bar b$ process, the extraction 
and measurement of the signals of these decay modes requires suppression of 
the $b \bar b$ contribution by a factor of twenty or better while preserving a 
good efficiency. 

The measurement of the top Yukawa coupling as well as the 
top-quark mass measurement will require efficient $b$-tagging to reduce 
combinatorial background in the reconstruction of the six and eight jet final
states. 
If Supersymmetry is realized in nature, the study of its rich Higgs sector 
will also require efficient identification of $b$-jets and $\tau$ leptons to 
isolate the signals for the decays of the heavier $A^0$, $H^0$ and $H^{\pm}$ 
bosons from the severe combinatorial backgrounds in the resulting complex 
multi-jet hadronic final states. 
Due to the large expected $b$-jet multiplicity, highly 
efficient tagging is required to preserve a sizeable statistics of the 
signal events. Finally, both $b$ and $c$-tagging will be important in the study
of the quark scalar partners, while $\tau$ identification may be 
instrumental in isolating signals from Gauge Mediated Supersymmetry 
Breaking. 

Due to the large boost in typical 4-jet events at $\sqrt{s}$ = 0.5~TeV, 
short-lived $B$, $D$ and $\tau$ particles have large decay distances. 
These are significantly larger compared to those at LEP/SLC, the 
average boost of $B$ hadrons increasing from $\simeq 6$ to $\simeq 20$. 
An useful figure of merit of the ability to
identify secondary particles is the resolution on the impact parameter
$\sigma_{IP}$, defined as the distance of closest approach of the extrapolated 
particle trajectory to the $e^+e^-$ annihilation point. This resolution can be 
parametrised as the convolution of an asymptotic term $\sigma_{asymptotic}$, 
depending on the sensor single point resolution and the detector geometry, with
a multiple scattering term $\sigma_{ms}$, depending on the thickness of the 
sensor and of its support structure:
$\sigma_{I.P.}^{R-\phi (R-z)} = \sigma_{asymptotic} \oplus \frac{\sigma_{m.s.}}
{p~\sin^{3/2 (5/2)}\theta}$ where $p$ is the particle momentum and $\theta$ 
its polar angle. In Table~1 the average charged decay multiplicity, 
decay distance in space, and track impact parameter are summarised.
\begin{table}[h!]
\captive{The average decay charged multiplicity, decay distance in space and 
track impact parameter for $B$, $D$ and $\tau$ decays in four jet events at
$\sqrt{s}$ = 500~GeV
\label{tab:1}}
\begin{center}
\begin{tabular}{|l|c|c|c|}
\hline
Decay & $<N_{sec}>$ & $<d_{space}>$ (cm) & $<IP_{R-\phi}>$ ($\mu$m) \\
\hline \hline
$B \rightarrow X$          & 4.9 & 0.8 & 450 \\
$D \rightarrow X$          & 2.3 & 0.4 & 150 \\
$\tau^{\pm} \rightarrow X$ & 1.3 & 0.1 & ~40 \\ 
\hline
\end{tabular}
\end{center}
\end{table}
Heavy particle identification relies on the accurate extrapolation of the 
particle trajectories to their production point to identify secondary decay
particles from those produced at the primary vertex in the hadronisation 
process. Efficient tagging of short-lived $\tau$ leptons requires 
very accurate impact parameter resolution for a single high momentum particle 
in order to discriminate single-prong $\tau$ decays from electrons and muons.
From the typical impact parameters given in Table~1, a resolution 
$\sigma_{asymptotic}< 10$~$\mu$m is required. 
On the contrary the decay multiplicity of $B$ and $D$ particles is large 
enough to perform inclusive reconstruction of decay vertices and efficient 
$b/c$ flavour separation relies on the distinctive differences 
in secondary vertex topology, invariant mass and charged decay multiplicity 
between beauty and charm hadrons. In order to fully exploit these features, 
sensitivity to $1~\mathrm{GeV/c} < p < 3.5~\mathrm{GeV/c}$ particles 
originating at secondary or tertiary vertices must be retained. From the 
typical impact parameter difference for $B$ and $D$ decay products, requiring 
$> 3~\sigma_{IP}$ separation for the majority of the particles implies value 
of $\sigma_{ms} < 30$~$\mu$m. As an exemplification of these requirements, the
performance of a charm-tagging algorithm is shown in Figure~1
for two assumptions on the impact parameter resolution.
\begin{figure}
\begin{center}
\epsfig{file=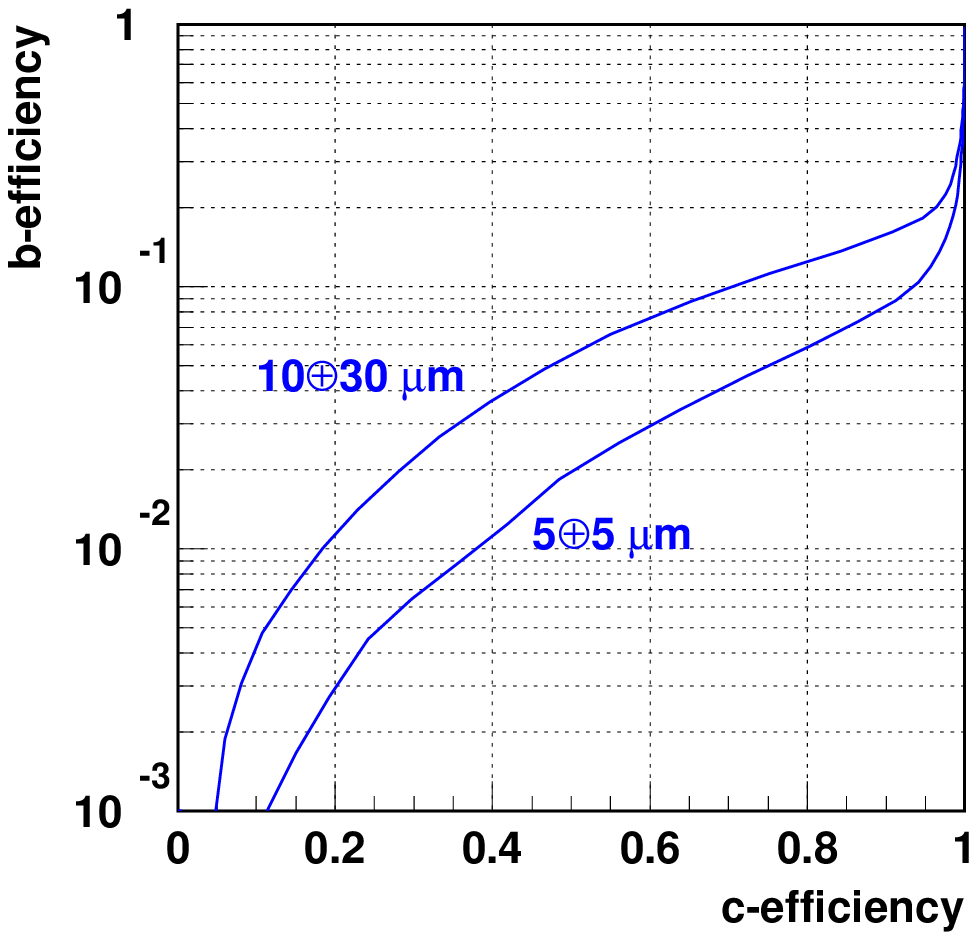,height=8.0cm}
\end{center}
\captive{The $b$-quark background efficiency as a function of the $c$-quark 
signal efficiency for a $c$-tag algorithm based on topological vertexing and 
impact parameters, developed on the basis of the experience from the SLC and 
LEP experiments. The two curves show the expected performance for two different
assumptions on the track impact parameter resolution.}
\label{fig:ebec}
\end{figure}
Due to the forward enhanced cross-section of several processes of interest,
such as $e^+e^- \rightarrow WW \rightarrow H^0 \nu \bar \nu$, and of the 
$\gamma \gamma$ background, the Vertex Tracker must provide tagging 
capabilities for polar angles down to $\cos \theta \simeq 0.95$.
Lepton charge determination in $e^+e^- \rightarrow W^+W^- 
\rightarrow X \ell \bar \nu$ decays and luminosity measurement by large angle 
Bhabha scattering require the extension of tracking coverage down to 
$\cos \theta \simeq 0.99$.

These motivations have promoted a significant amount of studies on the Vertex 
Tracker conceptual design, the choice of the detector technology and the 
definition of the needs of dedicated R\&D and the evaluation of its physics 
performances. 
Contrary to most of the other detector components, such as the main tracking 
chamber or the calorimeters for which the linear collider specifications can 
be addressed with technologies already developed for the LEP/SLC and LHC 
experiments, the Vertex Tracker sets very challenging 
requirements that can only be met by a detector of new generation.  
Therefore, an active programme of studies is presently carried out by US, 
Asian and European groups defining the design of the detectors for a future 
linear collider~\cite{sitges}.
In this paper we review the present status of the conceptual design, detector 
R\&D and physics performances of the Vertex Tracker design for the TESLA
$e^+e^-$ collider project. The next chapter presents the predicted 
experimental conditions at the TESLA interaction region. In chapter~3 the 
proposed solutions for the silicon sensor technology and the tracker 
conceptual designs are reviewed. The last chapter summarises the estimated 
performances and their impact on Higgs boson physics.

\section{Backgrounds and Physics events\\ at the TESLA $e^+ e^-$ collider}

The TESLA project~\cite{CDR} is based on the use of superconducting 
accelerating structures that deliver very long beam pulses ($\sim$ 900 $\mu$s),
each accelerating a train of up to 4500 bunches. This scheme allows a large 
bunch spacing (190-340~ns) making it possible, for the detector, to 
resolve single bunch crossings (BX) and to perform fast bunch-to-bunch 
feedback needed to stabilise the beam trajectory within a train, thus 
preserving the nominal luminosity of 
(3 - 5)~$\times 10^{34}$~cm$^{-2}$~s$^{-1}$. 
The large luminosity of each individual bunch-crossing and the large number
of bunches in a single pulse imply a high rate of background events that needs 
to be minimised by a Vertex Tracker with very high granularity and fast time 
stamping to identify the bunch corresponding to the physics event of interest. 
While there are several sources of backgrounds, the most relevant for this
discussion are the incoherent pair creation, neutrons and $\gamma \gamma$
backgrounds~\cite{schulte}.

Low energy $e^+e^-$ pairs are created in the e.m. interaction of the colliding
beams by the $\gamma \gamma \rightarrow e^+ e^-$ (Breit-Wheeler), 
$e^{\pm} \gamma \rightarrow e^{\pm} e^+ e^-$ (Bethe-Heitler) and
$e^+ e^- \rightarrow e^+ e^- e^+ e^-$ (Landau-Lifschitz) processes~\cite{pair}.
The generated electrons and positrons are confined spiralling at small radii 
by the solenoidal magnetic field of the experiment. The number of particles
from pairs, intercepting a surface at a given radius depends on the 
intrinsic transverse momentum of the pairs and on their deflection from the 
e.m. interaction with the opposite beam. Particles are thus confined in an 
envelope defined by the maximum angle of deflection and by the strength of the
magnetic field $B$. 
For the design of the Vertex Tracker the two main parameters
of interests are the maximum radius $R_{max}$ of the envelope and its 
longitudinal position of crossing $z_c$ with a 
reference surface located at radius $R$. These define the minimum radius and 
maximum length of the innermost detector layer and can be expressed in terms 
of the simple scaling laws~\cite{hawaii}:
\begin{eqnarray}
R_{max} [cm] = 0.35 \sqrt{\frac{N}{10^{10}} \frac{1}{B [Tesla]} z [cm] 
\frac {1}{\sigma_z [mm]}}
\end{eqnarray}
\begin{eqnarray}
z_c [cm] = 8.3~ R^2 B [Tesla] \sigma_z [mm] \frac{10^{10}}{N} .
\end{eqnarray}
where $N$ is the number of particles in one bunch, $B$ is the magnetic
field strength and $\sigma_z$ the bunch length. Detailed simulation for the 
case of the TESLA collider at $\sqrt{s}$ = 500~GeV and with $B$ = 3~T, shows 
that the density of hits generated by pairs at a radius of 1.2~cm from the 
interaction point corresponds to 
$\simeq$ 0.2 hit mm$^{-2}$ BX$^{-1}$ and that the longitudinal positions of
the pair envelope crossing point on a scoring cylinder at this radius are at
$\pm 7$~cm. In addition to the pairs, dense hadronic jets in high multiplicity
events, such as $e^+e^- \rightarrow t \bar t$; $t \rightarrow W b$, 
$W \rightarrow q \bar q'$ contribute to the detector occupancy. It has been 
calculated that at $R$ = 3.0~cm about 20\% (10\%) of the particles in a jet
have at least one additional hit within a 150~$\mu$m distance in the 
$R-\phi$ ($R-z$) plane. These results rule out the application of silicon 
microstrip detectors and highlight the importance of detectors with small 
sensitive cells, such as pixel sensors.
\begin{figure}
\begin{center}
\epsfig{file=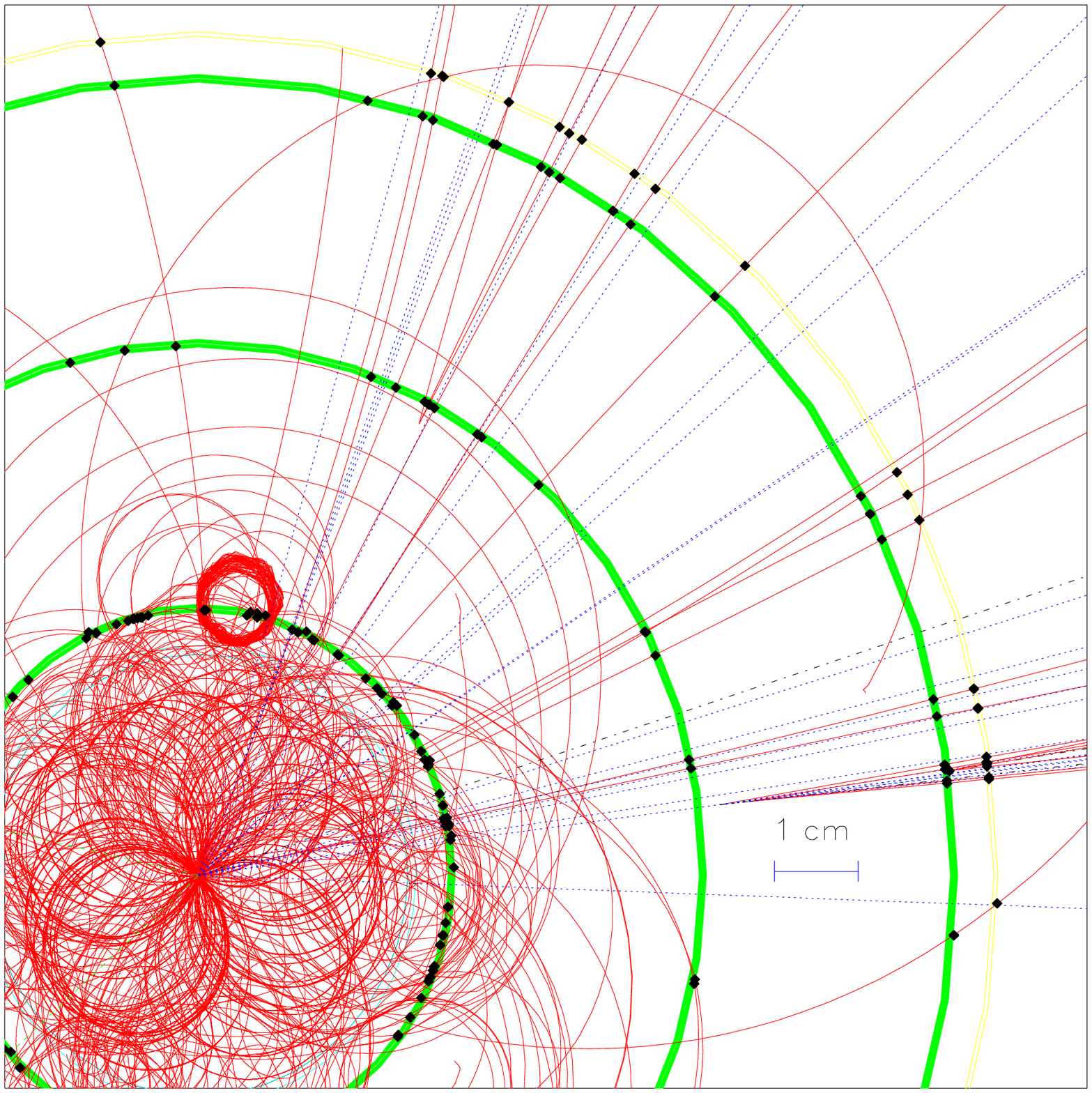,width=7.5cm}
\end{center}
\captive{A simulated $e^+e^- \rightarrow Z^0 H^0$; 
$Z^0 \rightarrow \mu^+ \mu^-$, $H^0 \rightarrow b \bar b$ event at $\sqrt{s}$
= 500~GeV overlayed to the pair backgrounds. The high track density and the 
distinctive structure of secondary vertices from the $B$ hadron decays are 
clearly visible. It is interesting to point out that 20\% of the $B$ hadrons 
in such events will decay after having traversed the innermost Vertex Tracker
layer.}
\label{fig:bb}
\end{figure}

Another source of background to be taken into account in the choice of the 
detector technology is the flux of neutrons photo-produced at the dump of 
electrons from pairs and radiative Bhabha scattering and of beamstrahlung 
photons. Photons in electro-magnetic showers, lead to the release of fluxes 
of neutrons through giant dipole resonance excitation, pseudo-deuteron 
mechanism and photo-pion reaction. 
Neutrons may induce permanent radiation damage to semiconductor devices by
nuclear interactions generating displacement damage of the bulk material. 
Bulk damage results in charge collection and charge transfer inefficiency 
(CTI) in the detector structure. CTI affects in 
particular devices such as the CCDs, where the ionisation charge typically 
undergoes several thousands transfers before reaching the output node. 
The computation of the neutron flux at the 
interaction region relies on the modelling of their production and transport 
in the accelerator tunnel and in the detector and it is still subject to 
significant uncertainties. Estimated fluxes are of the order of a few 
10$^9$ $n$~(1~MeV)~cm$^{-2}$~year$^{-1}$~\cite{Tesch,Ye}, where the 
anticipated neutron flux has been normalised in terms of equivalent
1~MeV neutrons assuming NIEL scaling.

Finally the cross-section for two photon events, at the linear collider 
energies, is many orders of magnitude larger compared to that of 
$e^+e^-$ annihilation events. Due to the large luminosity per bunch crossing
obtained by TESLA,
$L = 2.2~\times~10^{-3}$~nb$^{-1}$~BX$^{-1}$, the probability for a 
$\gamma \gamma \rightarrow jets$ event overlapping with a physics event is
0.09 - 0.14~$\times~n_{BX}$, where $n_{BX}$ is the number of bunch crossings 
integrated in a detector read-out cycle. Two photon events, depositing
a significant energy in the forward regions, can seriously interfere with the 
identification of physics processes whose signature is missing energy such
as $e^+e^- \rightarrow H^0 \nu \bar \nu$ and SUSY decays and result in an 
important source of systematics for cross-section measurements. Therefore 
single bunch identification is highly desirable in order to minimise this
background.  
 
\section{Detector technologies and\\ Vertex Tracker conceptual designs} 

The linear collider Vertex Tracker must be able to provide
a track impact parameter resolution better than $10~\mu m \oplus 
\frac{30~\mu m~{\mathrm GeV/c}}{{\mathrm p}~\sin^{3/2}\theta}$ in both 
the $R-\phi$ and $R-z$ projections for jet flavour identification, have 
sensitive cells of 150 $\times$ 150~$\mu$m$^2$ or less to keep the occupancy 
from pairs and hadronic jets below 1\% and possibly identify 
single bunch crossings separated by about 200~ns to minimise pair and 
$\gamma \gamma$ backgrounds. 
There are two types of such silicon sensors, already used at
collider experiments, that have the potential to satisfy these specifications 
in terms of sensitive cell size and have been already proposed for this 
application: Hybrid Pixels sensors~\cite{hawaii,yamamoto,MBa}, i.e. pixel 
detectors bump-bonded to their VLSI read-out chip, and Charged Coupled Devices 
(CCD)~\cite{ccdlc}. In addition Monolithic CMOS sensors have also been 
recently considered~\cite{yamamoto,MCa}.

\subsection{Hybrid and Monolithic Pixel detectors}

\begin{figure}[hb!]
\begin{center}
\begin{tabular}{c c}
\epsfig{file=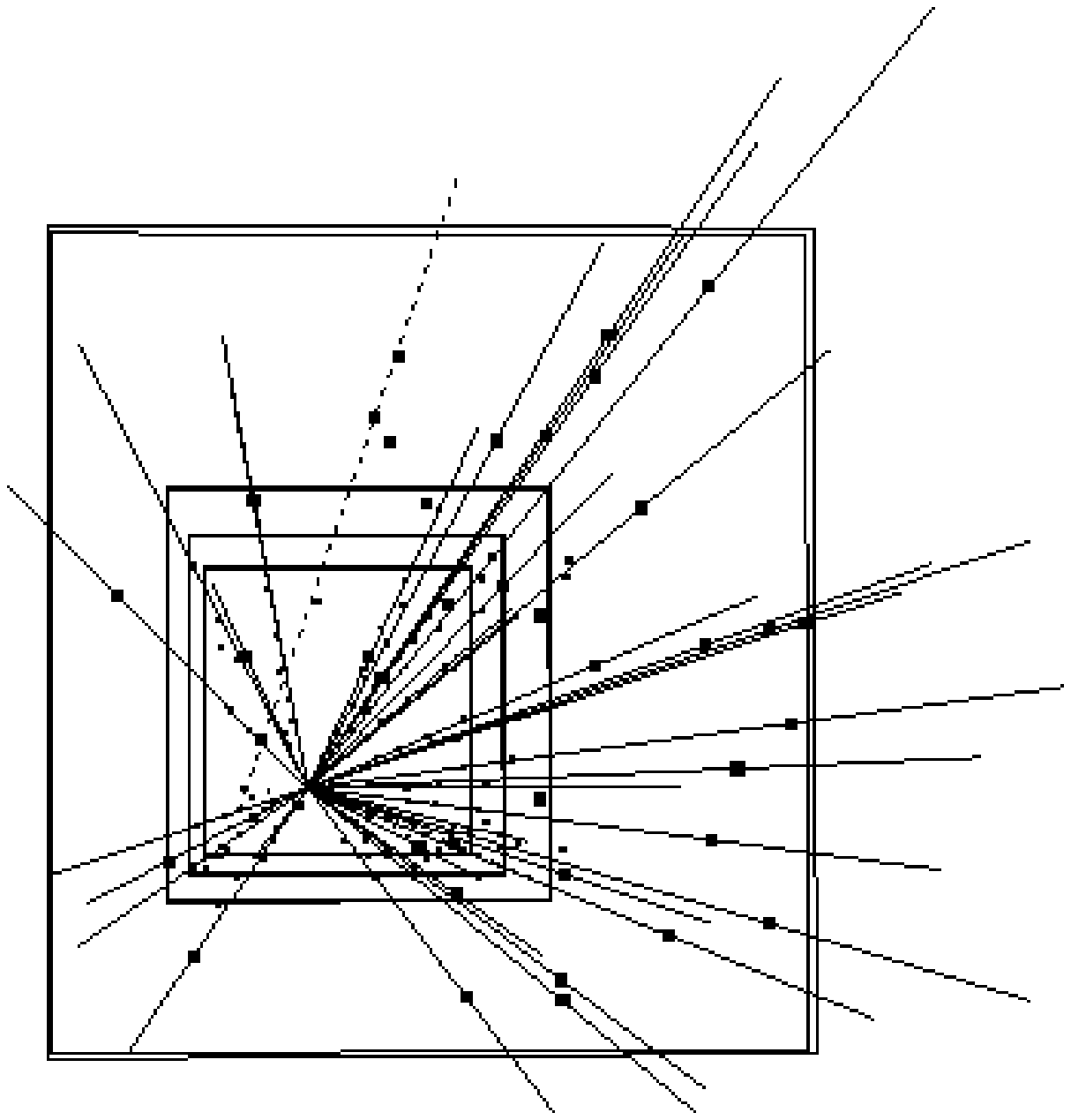,bbllx=133,bblly=215,bburx=478,bbury=576,
width=6.75cm,clip=} &
\epsfig{file=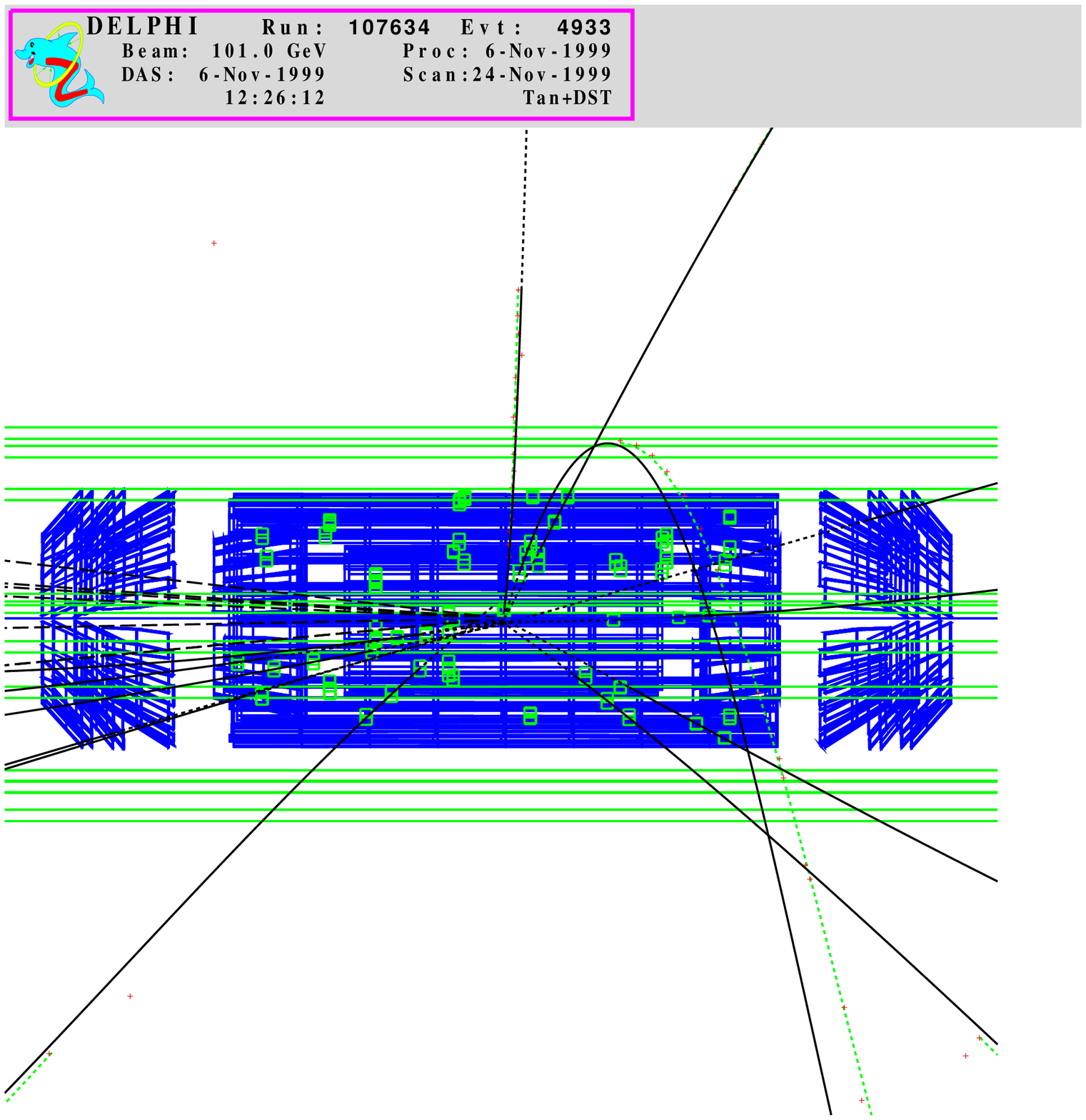,bbllx=15,bblly=180,bburx=525,bbury=610,width=8.5cm,
clip=}
\end{tabular}
\end{center}
\captive{Two examples of applications of hybrid pixels sensors: an
event of Pb-Pb collision recorded by the WA97 pixel telescope~\cite{wa97} 
(left) and an hadronic event with tracks reconstructed by the DELPHI VFT in a
LEP $e^+e^-$ annihilation at $\sqrt{s}$~=~204~GeV (right).} 
\label{fig:pixe}
\end{figure}
Pixel detectors are the natural successors of the microstrip detector design 
widely used in high energy physics experiments at fixed target and colliders.
By further segmenting the strip implant, truly three-dimensional reconstruction
of the point of impact of the ionising particle can be obtained. The read-out
is performed by a bump-bonded VLSI chip in hybrid detectors or by integrating
the circuitry on the same wafer in the monolithic design.

Hybrid silicon pixel detectors have been developed and successfully applied to 
track reconstruction in high energy physics experiments in the last decade.
In particular, {\sc Delphi} at LEP was the first collaboration 
adopting hybrid pixel sensors for a Vertex Tracker at a collider 
experiment~\cite{SiTracker}. 
The Very Forward Tracker included four crowns of 330 $\times$ 330~$\mu$m$^2$
pixel sensors bump-bonded to read-out VLSI modules for a total of 1.2M
channels. The fraction of noisy masked pixels was $\simeq$ 0.3\% and the
detector efficiency of the order of 95\%. Hybrid pixel sensors have been 
further developed for ALICE~\cite{alice}, ATLAS~\cite{atlas} and 
CMS~\cite{cms} to meet the experimental conditions of the LHC collider. 
These R\&D activities have demonstrated the feasibility of fast time stamping 
(25~ns) and sparse data scan read-out, and the operability of hybrid pixel 
detectors exposed to neutron fluxes well beyond those expected at the linear 
collider. 

The linear collider application now defines new areas of specific R\&D 
aimed to improve the achievable single point resolution to better than
$10~\mu$m and to reduce the sensor+VLSI thickness. 
The detector resolution requirement can be accomplished by sampling the 
diffusion of the charge carriers generated along the particle path and 
assuming an analog read-out to interpolate the signals of neighbouring cells.
Given that the charge diffusion is 
$\sim {\rm 8~\mu m}$ in ${\rm 300~\mu m}$ thick silicon, its efficient 
sampling and signal interpolation requires a pitch of not more than 25~$\mu$m.
This has been successfully proven to work in one-dimensional microstrip 
sensors \cite{Anna}.
\begin{figure}[hb!]
\begin{center}
\epsfig{file=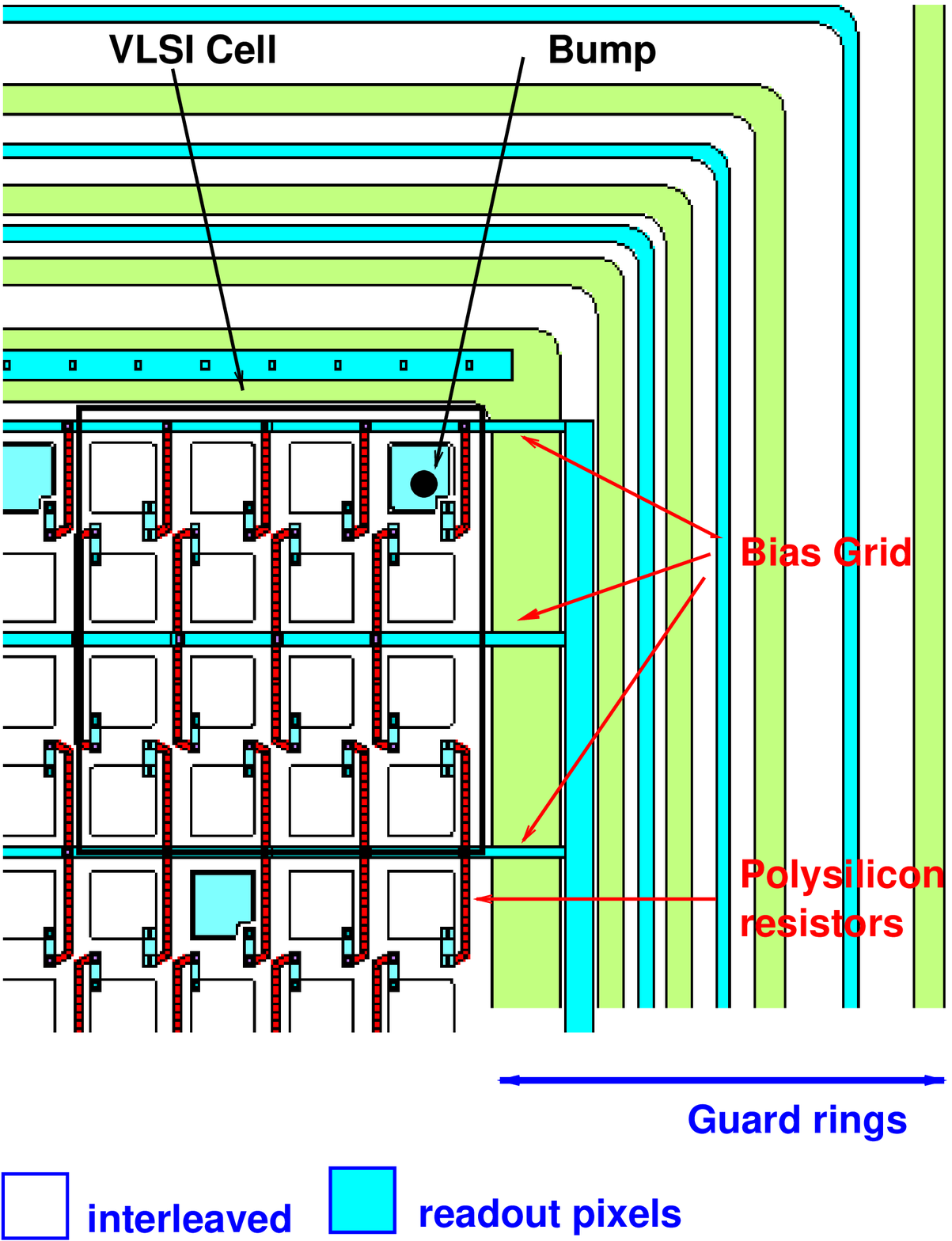,height=7.5cm}
\captive{Layout of the upper corner of pixel detector test structure, 
with 50~$\mu$m implant and 200~$\mu$m read-out pitch.}
\label{fig:corner}
\end{center}
\end{figure}
In pixel devices the ultimate read-out pitch is 
constrained by the front-end electronics, to be integrated in a cell matching 
the sensor pattern. At present, the most advanced read-out electronics have a 
minimum cell dimension of ${\rm 50 \times 300~\mu m^{2}}$ not suitable for a 
finely segmented charge sampling. 
The trend of the VLSI development and recent studies
\cite{Snow} on intrinsic radiation hardness of deep sub-micron CMOS technology 
certainly allows to envisage of a sizeable reduction in the cell dimensions 
on a linear collider time scale but sensor designs without such basic 
limitations are definitely worth being explored. 
A possible solution is to exploit the capacitive coupling of neighbouring 
pixels and to have a read-out pitch n times larger than the implant 
pitch~\cite{Bonvicini}.
The proposed sensor layout is shown in 
Figure~4 for n=4. In this configuration, the charge carriers 
created underneath an interleaved pixel will induce a signal on the 
capacitively coupled read-out nodes.
In a simplified model, where the sensor is reduced to a capacitive network, 
the ratio of the signal amplitudes on the read-out nodes at the left- 
and right-hand side of the interleaved pixel in both dimensions will be
correlated to the particle position and the resolution is expected to be 
better than ${\rm (implant~pitch)/\sqrt{12}}$ for an implant pitch of 
25~$\mu$m or smaller. The ratio between the 
inter-pixel capacitance and the pixel capacitance to backplane will play a 
crucial role, as it defines the signal amplitude reduction at the output
nodes and therefore the maximum number of interleaved pixels.
Calculations with such capacitive network models \cite{Pindo} show that 
resolutions similar to those achieved by reading out all pixels are
obtainable if the signal amplitude loss to the backplane is small.
Recent tests on a microstrip sensor, with 200~$\mu$m read-out pitch, 
have achieved a ${\rm 10~\mu m}$ resolution with three interleaved 
strip layout~\cite{Krammer}. Similar results are expected in a pixel
sensor, taking into account both the lower noise because of the
intrinsically smaller load capacitance and the charge sharing
in two dimensions. 
Reducing the read-out density, without compromising the achievable space
resolution, is also beneficial to limit the power dissipation and the 
overall costs. 
In order to verify the feasibility of this scheme a dedicated R\&D program has
started~\cite{hmk}. 
A prototype set of sensors with interleaved pixels and different 
choices of implant and read-out pitch have been already designed, produced 
and tested\cite{AZ,MCa}. 
These studies will determine the achievable single point resolution for 
hybrid pixel sensors.

Monolithic pixel sensors, integrating the read-out circuit on the same silicon
wafer acting as particle detector, have been proposed for visible 
imaging~\cite{cmos1} since the beginning of the decade. 
\begin{figure}[ht!]
\begin{center}
\epsfig{file=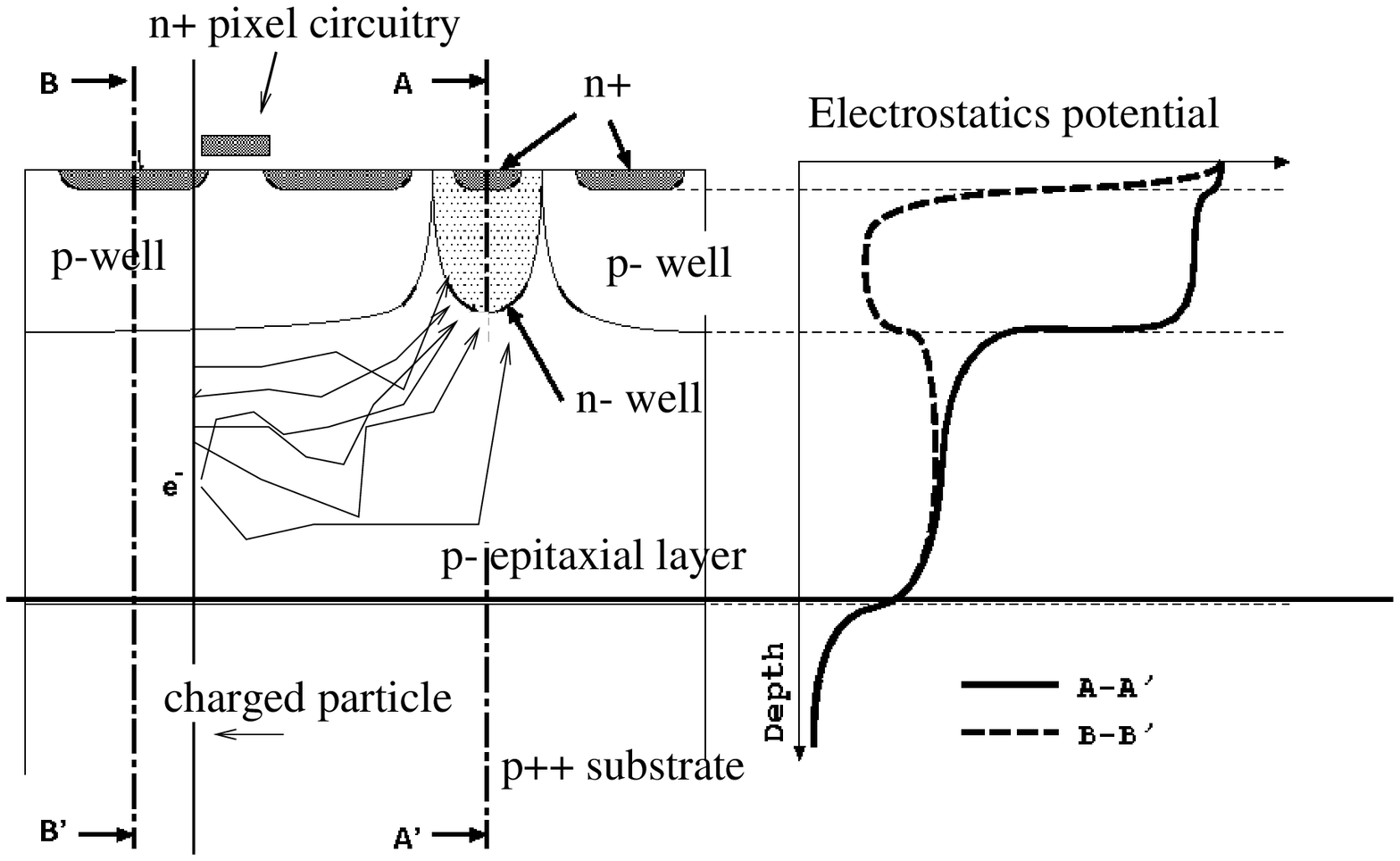,height=6.0cm}
\captive{The proposed monolithic pixel structure. The circuitry is integrated 
in the p-well while the photosite is a n-well diode on the p-epitaxial layer. 
Because of the difference in doping level, the p-well and the p++ substrate 
act as reflective barriers. The generated electrons are collected in the 
n-well.}
\label{fig:cmos2}
\end{center}
\end{figure}
These devices have several potential advantages including their low cost, 
inherent radiation hardness, possibility to integrate several functionalities 
on the sensor substrate including random access and very low power consumption
since the circuitry in each pixel is only active during read-out time. 
A limiting factor encountered in visible light applications is the low fill
factor, i.e. the fraction of the pixel area that is sensitive to the light.
In order to solve this problem for detection of ionising particles, it has
recently been proposed to integrate a sensor in a twin-well technology with a 
n-well/p-substrate diode (see Figure~5)~\cite{turch}. 
This technique has already proven its effectiveness in visible light 
applications \cite{cmos2} reducing the blind area to that of the metal lines, 
opaque for visible light but not for charged particles. Monolithic
CMOS sensors could achieve a high spatial resolution, the achievable pixel 
pitch being of the order of ${\rm 10~\mu m}$ and hence the spatial resolution 
better than ${\rm 3~\mu m}$ already with a binary read-out. At the same time, 
very low multiple scattering is introduced as the substrate can in principle 
be thinned down to a few microns. 
In order to prove the effectiveness of the CMOS sensor technique for the 
linear collider application, a R\&D program has been initiated~\cite{strasb}. 
Existing commercial devices are currently under test and the prototype of
a full-custom design sensor is being produced in ${\rm 0.6~\mu m}$ CMOS 
technology. 

\subsection{CCD detectors}

Charge coupled devices (CCD) have been successfully used in high resolution 
tracking detectors both at fixed target and collider experiments since the
mid 80's~\cite{damerell1}. The first application of CCDs at a collider 
experiment has been with the VXD1 in the SLD experiment at the SLC collider,
later followed by the VXD2 and VXD3 upgrades (see Figure~6). 
The VXD3 detector consisted of three cylindrical layers, for a total of 
307M 20 $\times$ 20 $\mu$m$^2$ pixels providing a space point resolution of 
3.8~$\mu$m~\cite{vxd3}. 
The read-out time of 180~ms caused the occasional recording of 
overlapping events. However, due to the low SLC duty cycle and luminosity and 
the low track density, this did not cause any problems to the track pattern 
recognition.
\begin{figure}[ht!]
\begin{center}
\begin{tabular}{c c}
\epsfig{file=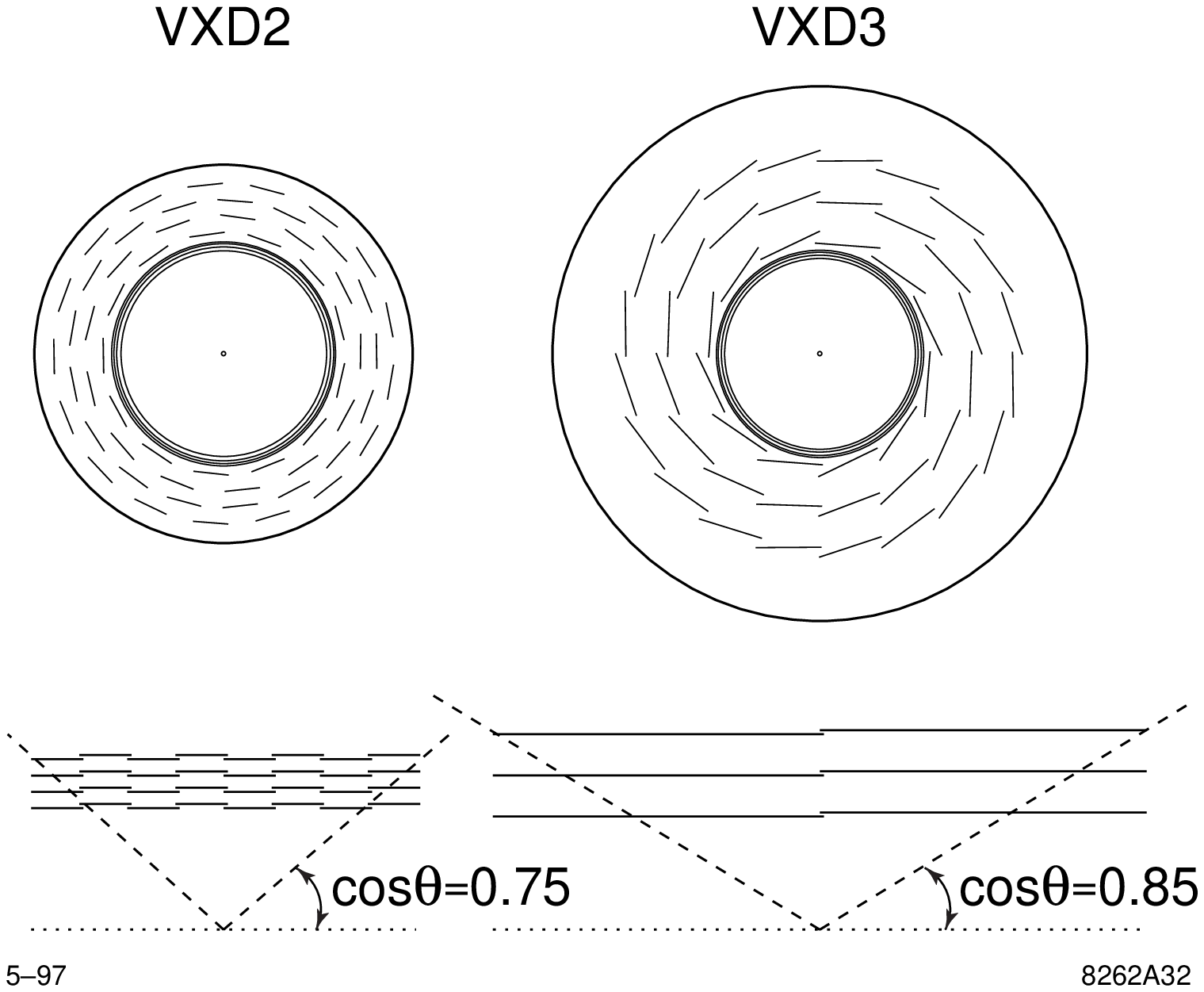,width=6.5cm} &
\epsfig{file=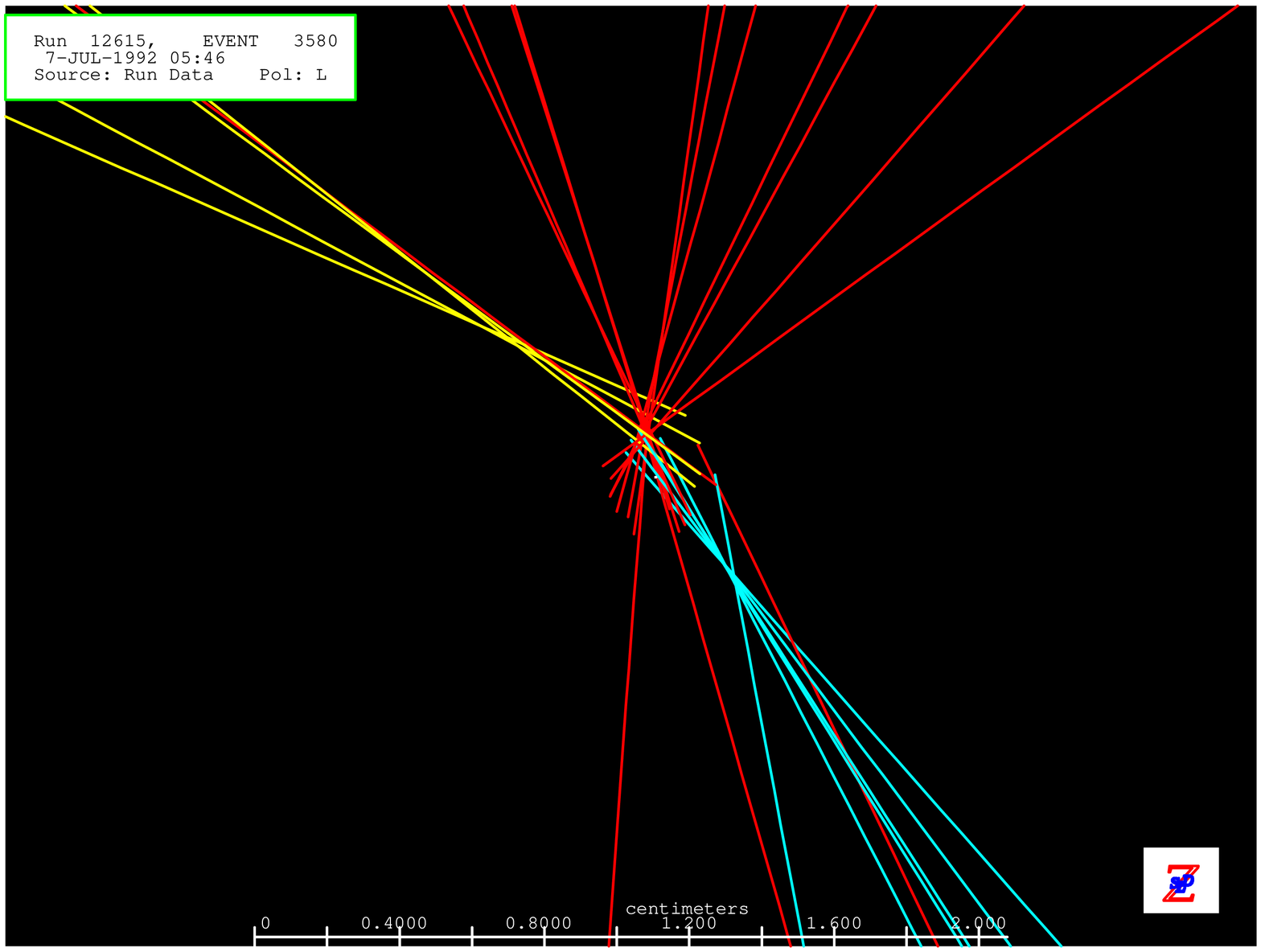,width=7.5cm}
\end{tabular}
\end{center}
\captive{Schematic layouts of the SLD VXD detectors and a magnified view of
a candidate $e^+e^- \rightarrow Z^0 \rightarrow b \bar b g$ decay with clearly
defined secondary vertices.} 
\label{fig:vxd3}
\end{figure}

For the linear collider application, CCD of larger area (up to 3000~mm$^2$)
and lower thickness (in principle only limited by the epitaxial layer 
thickness of $\simeq$~20~$\mu$m) are envisaged~\cite{CCDnew}. 
Developments in low noise electronics
can further improve the single point resolution to $\simeq$ 3~$\mu$m.
While CCD sensors have ideal characteristics in terms of the spatial 
resolution and detector thickness, they presently lack the required read-out 
speed necessary to cope with the TESLA bunch timing and are possibly sensitive
to neutron damage at fluxes of the order of that expected at the linear 
collider. An intense R\&D program is presently underway to overcome these 
limitations~\cite{ukfi,CCDnew}.
At the read-out speed of 5~MHz, adopted at SLD, the CCD detectors would
integrate a full TESLA train of 2850 (4500) bunches at $\sqrt{s}$ = 500 
(800)~GeV corresponding to 560 (900) hits mm$^{-2}$ and an occupancy
of 22 (36)\% on the innermost layer. These rates give an unacceptably high
number of hit association ambiguities and fake tracks. Even increasing the 
read-out speed to 50~MHz would not yet provide with a tolerable detector 
occupancy.
In order to solve this problem a new read-out scheme, named ``column 
parallel read-out'', has been proposed. In this scheme the serial read-out 
circuit is replaced by a read-out IC, with circuits every columns, bump-bonded
at the end of the CCD ladder (see Figure~7).
\begin{figure}[hb!]
\begin{center}
\begin{tabular}{c c}
\epsfig{file=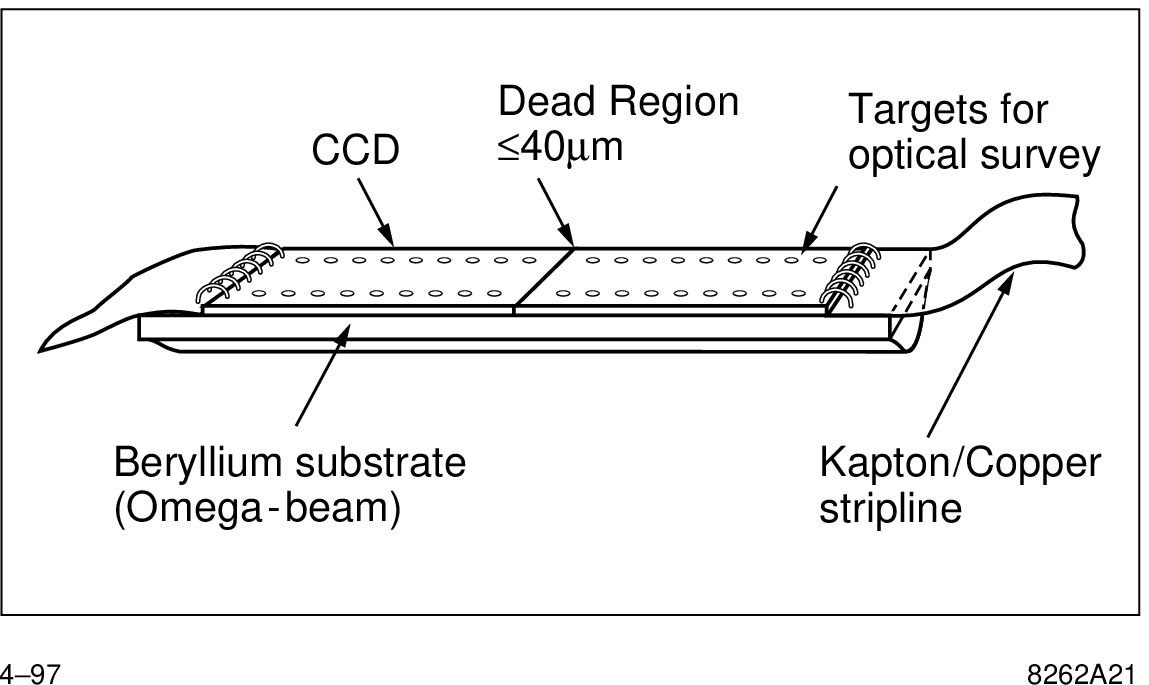,width=7.5cm} &
\epsfig{file=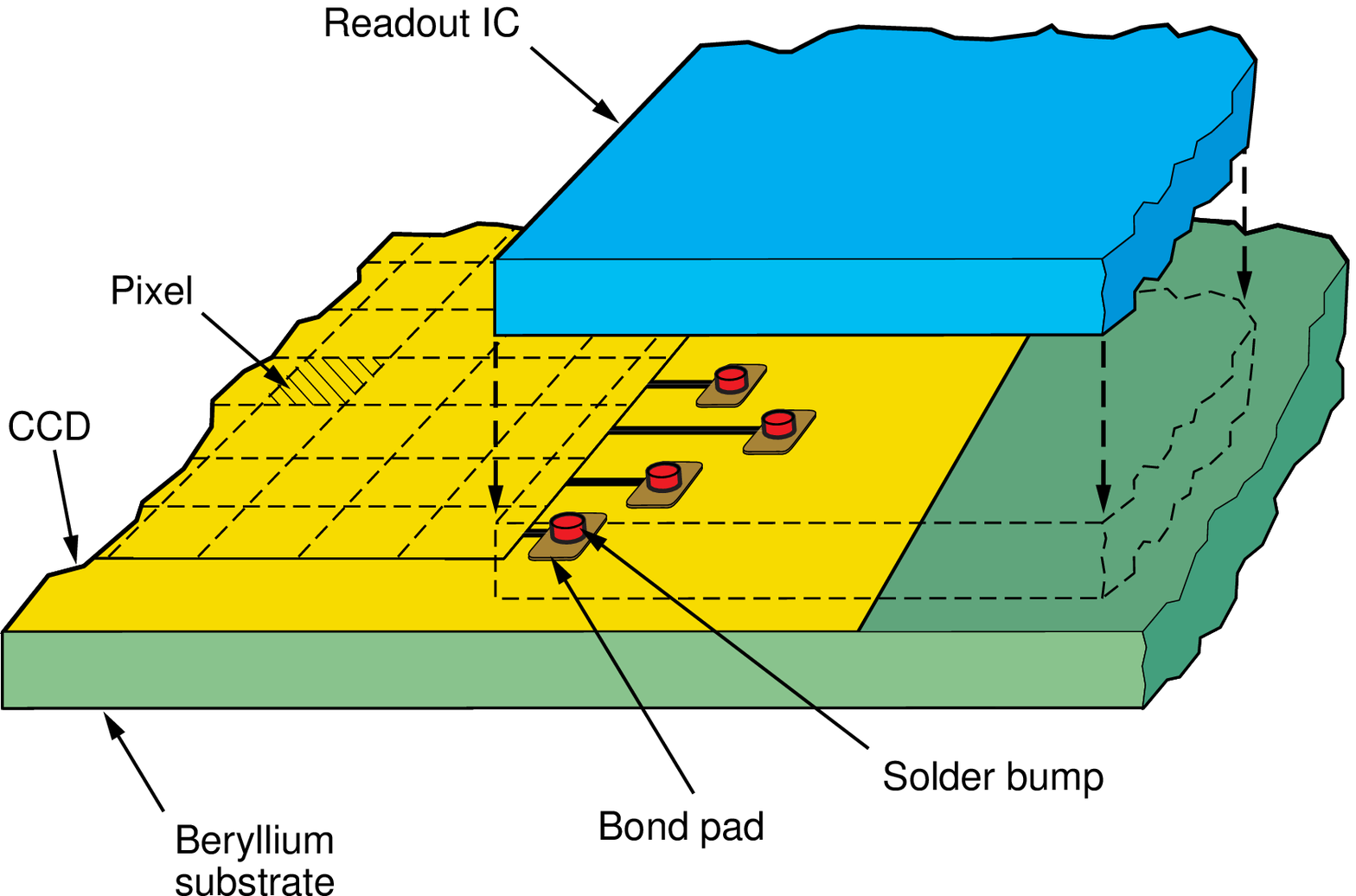,width=7.5cm} \\
\end{tabular}
\end{center}
\captive{The proposed layout of a CCD ladder (left) with read-out parallel
column read-out electronics bump-bonded at one end of the sensitive area 
(right).}
\label{fig:ladder}
\end{figure}
  
The column parallel read-out is aimed at reducing the number of integrated 
bunches to 60-100, thus achieving an acceptable density of background hits
at the innermost layer. There are further issues deriving from the increased 
power dissipation and the feasibility to drive CCDs at the proposed frequency
that need to be investigated. Also the effect of the high rate of 
$\gamma \gamma$ overlap to physics events, consequent to the still long 
integration time of CCD detectors, has still to be studied in details.   
The CCD sensitivity to radiation damage also requires a careful evaluation.
As mentioned above, neutron induced displacements in the bulk silicon
create trap regions that reduce the charge transfer efficiency. Due to the 
several hundreds or thousands of transfers the ionisation charge undergoes 
before the read-out node, the charge transfer inefficiency (CTI) must be kept 
below $\simeq 10^{-4}$ not to significantly deteriorate the sensor efficiency. 
Recent tests of CCD detectors have shown that detector inefficiencies in 
excess of $\simeq$ 10\% are observed after fluxes of several times 
10$^9$ $n$ (1~MeV) cm$^{-2}$, unless the detector is cooled to 
$T < 185^{\circ}K$~\cite{ccdneutr}. 
In fact, at low enough temperature the lifetime of trapped 
electrons becomes very long and the traps remain filled during the transfer of 
subsequent electron packets. In addition, the column parallel read-out scheme,
decreasing the number of charge transfers before the read-out node, eases the 
requirements for the maximum tolerable CTI. The radiation resistance of CCD 
sensors represents a crucial issue in the choice of the detector technology
to be reviewed as more experimental data on neutron damage and updates of
the final focus region layout and improved neutron flux calculations will
become available. 

\subsection{Conceptual designs}

The basic geometrical structure of the Vertex Tracker consists of concentric 
cylinders of detectors complemented by crowns or disks in the forward regions.
In the TESLA design, the first detector layer, closest to the interaction 
region, is located at a radius of 1.2~cm outside a 0.5~mm thick Be beam-pipe. 
The beam-pipe is shaped to accommodate the envelope of deflected pairs
growing in radius moving away from the interaction region along the beam-line.
The distance of the first detector layer from the beam collision point is 
reduced by more than a factor of~2 compared to that at the SLC and by a factor
of~5 compared to LEP to the benefit of the achievable track extrapolation 
accuracy.
Since hybrid pixel detectors and CCDs have different performances in terms of
the space resolution and the requirements for mechanical support and cooling, 
the proposed tracker designs differ accordingly in several details.

\begin{figure}[h!]
\begin{center}
\begin{tabular}{|c|}
\hline
\epsfig{file=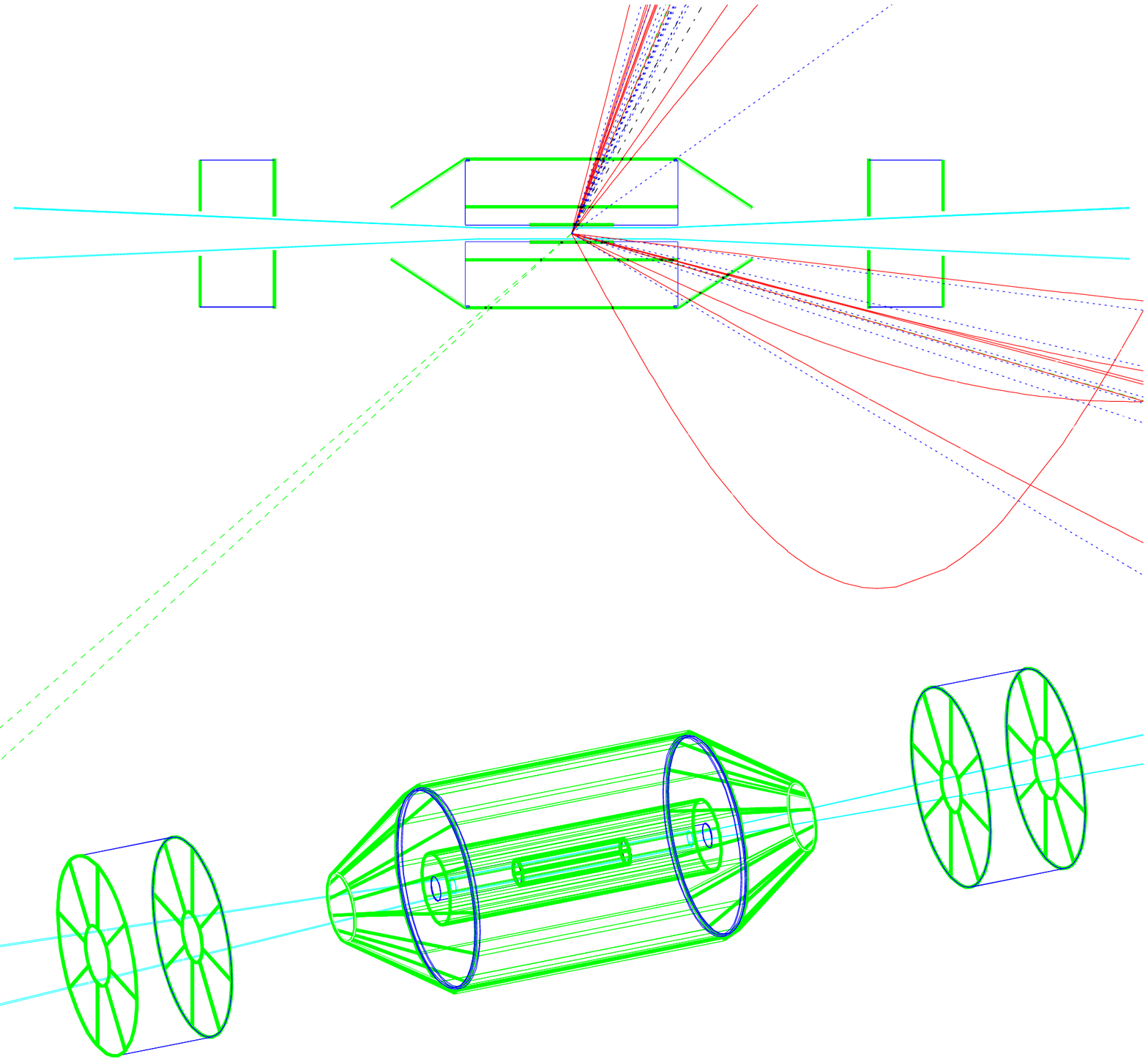,height=9.5cm}\\
\hline
\end{tabular}
\end{center}
\captive{The proposed layout of the Vertex Tracker with a simulated
$e^+e^- \rightarrow h^0 Z^0 \rightarrow b \bar b \mu^+ \mu^-$ event
overlayed. The three layered barrel
section is complemented by a crown and two disks of detectors to ensure 
accurate tracking in the forward region.}
\label{fig:layout} 
\end{figure}
The layout of the TESLA Vertex Detector~\cite{hawaii} based on hybrid
pixel sensors is shown in Figure~8 and consists of
a three-layer cylindrical detector surrounding the beam-pipe complemented
by forward crowns and disks extending the polar acceptance to small angles
following a solution successfully adopted in the DELPHI Silicon Tracker. 
The three barrel layers have a polar acceptance down to 
$|\cos \theta| = 0.82$. At lower angles, additional space points are obtained 
by extending the barrel section by a forward crown and two disks of detectors 
providing three hits down to $|\cos \theta| = 0.995$. The transition from the 
barrel cylindrical to the forward conical and planar geometries optimises the 
angle of incidence of the particles onto the detector modules in terms of the
achievable single point resolution and the multiple scattering contribution. 
Overlaps of neighbouring detector modules provide an useful mean of 
veryfing the relative detector alignment using particle tracks from dedicated 
calibration runs taken at $Z^0$ centre-of-mass energy. 
The requirement on the multiple scattering contribution to the track 
extrapolation resolution, lower than 30~$\mu$m/$p_t$, and the need to 
minimise the amount of material in front of the calorimeters and to ensure 
the optimal track matching with the main tracking system, set a constraint on 
the material budget of the Vertex Tracker to less than 3~\% of a radiation 
length ($X_0$).
These requirements can be fulfilled by adopting 200~$\mu$m thick detectors and 
back-thinning of the read-out chip to 50~$\mu$m, corresponding to 0.3~\%~$X_0$
of a radiation length, and a light support structure. The present concept for
the mechanical structure envisages the use of diamond-coated carbon fiber
detector support layers acting also as heat pipes to extract the heat 
dissipated by the read-out electronics uniformly distributed over the whole
active surface of the detector.
Assuming a power dissipation of 60~$\mu$W/channel, the total heat flux is 
450~W, corresponding to 1500~W/m$^2$, for a read-out pitch of 
200~$\mu$m. Preliminary results from a finite element analysis show that pipes 
circulating liquid coolant must be placed every 5~cm along the longitudinal 
coordinate except for the innermost layer where they can
be placed only at the detector ends to minimise the amount of material.
Signals can be routed along the beam pipe and the end-cap disks to the 
repeater electronics installed between the Vertex Tracker and the forward
mask protecting the Vertex Tracker from direct and backscattered radiation
from the accelerator. The estimated material budget corresponds to 
1.6\%~$X_0$ for the full tracker. 

\begin{figure}[h!]
\begin{center}
\epsfig{file=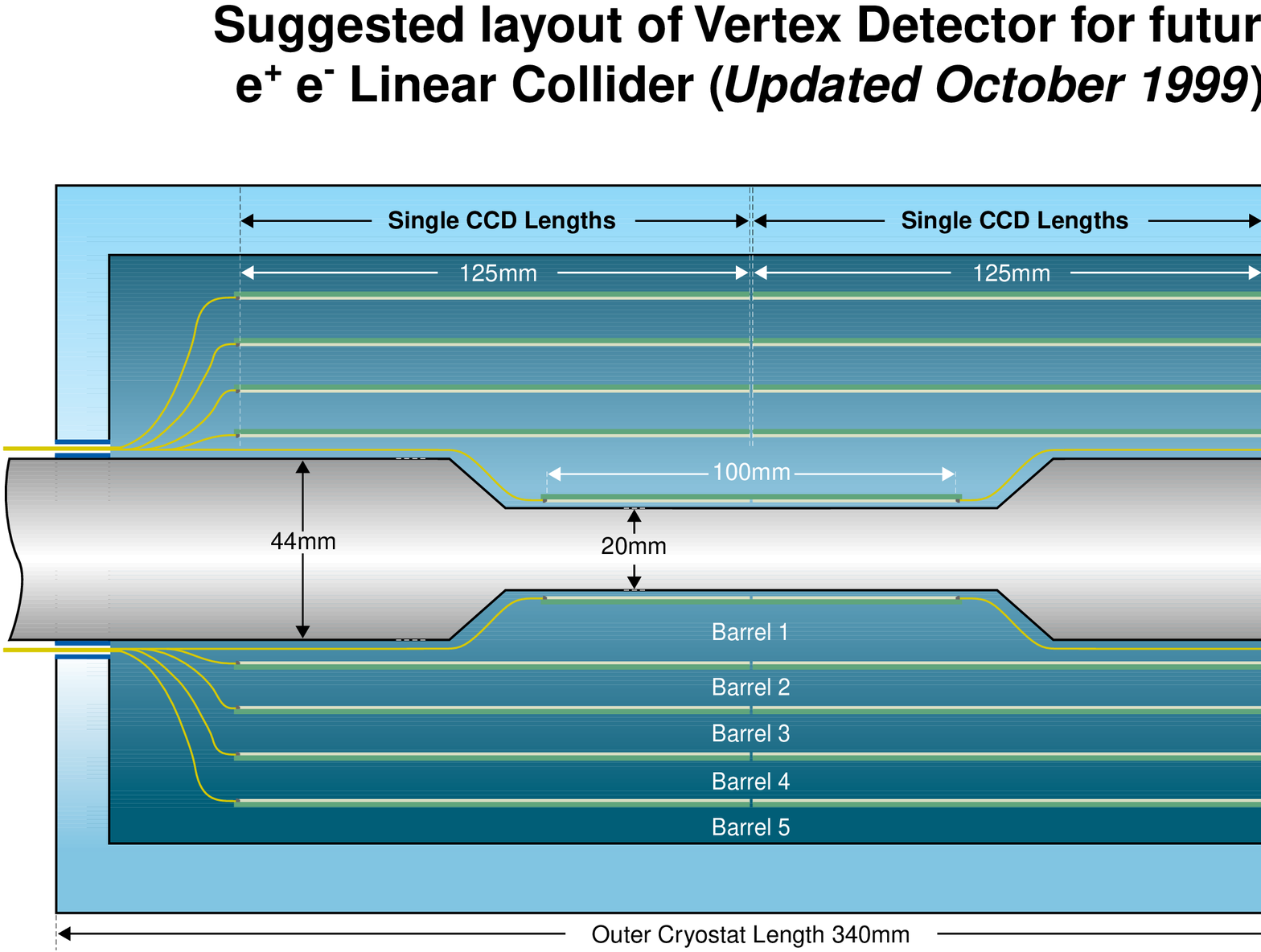,height=8.5cm,clip=}
\end{center}
\captive{The conceptual design of the CCD Vertex Tracker}
\label{fig:lc_ccd}
\end{figure}
The proposed CCD Vertex Tracker consists of five layers of CCD sensors
(see Figure~9) providing at least three hits down to $|\cos \theta| = 0.96$ 
and at least one down to $|\cos \theta|$ = 0.98. 
In the forward region, the track 
reconstruction relies on the match of the CCD hits with those recorded in 
disks of pixel detectors similarly to the previous design. 
The use of five layers of CCD sensors ensures self-consistent tracking 
capabilities with redundancy. This is possible due to the small multiple 
scattering contribution expected for thinned sensitive layers on a thin flat 
Be support structure corresponding to 0.11\%~$X_0$/layer. This approach has its
point of strength in the higher accuracy that can be achieved in the internal
Vertex Tracker alignment, due to the redundant track trajectory measurements,
compared to the relative alignment to the main tracking chamber. In 
addition, the external foam-insulated cryostat can be kept thin enough not to 
significantly interfere in the linking with the tracks extrapolated from the 
central tracker towards the vertex. 
The CCD thinning procedure has been studied in details and finite element 
analysis has shown that the deformations of the detectors mounted on a 
properly shaped Be support are within a tolerable range. 
The total material budget for the CCD Vertex Tracker has been estimated to be
0.70\%~$X_0$ plus the cryostat that, however, does not degrade to the track 
extrapolation accuracy achieved with the Vertex based track fit if the rate
of ambiguities is low enough. The light structure and the high accuracy of CCD
Vertex detector makes it the most ambitious and performant tracker presently 
under study.

\section{Vertex Tracker performances}

The geometry optimisation and the study of
the physics performances of the Tracker design has been performed with a 
GEANT based simulation of the detector, accounting for benchmark 
physics processes with enhanced forward production cross-section, such as 
$e^+e^- \rightarrow H^0 \nu \bar \nu$, pair and $\gamma \gamma$ backgrounds, 
local pattern recognition and detector inefficiencies. The impact parameter
resolution has been obtained by a Kalman filter based track fit to the 
associated hits in the Vertex Tracker and in the TPC.
The impact parameter resolutions, obtained for tracks with 
$|\cos \theta| < 0.92$, with the Hybrid Pixel and CCD Vertex Tracker options
are shown in Figure~10.
\begin{figure}[hb!]
\begin{center}
\begin{tabular}{c c}
\epsfig{file=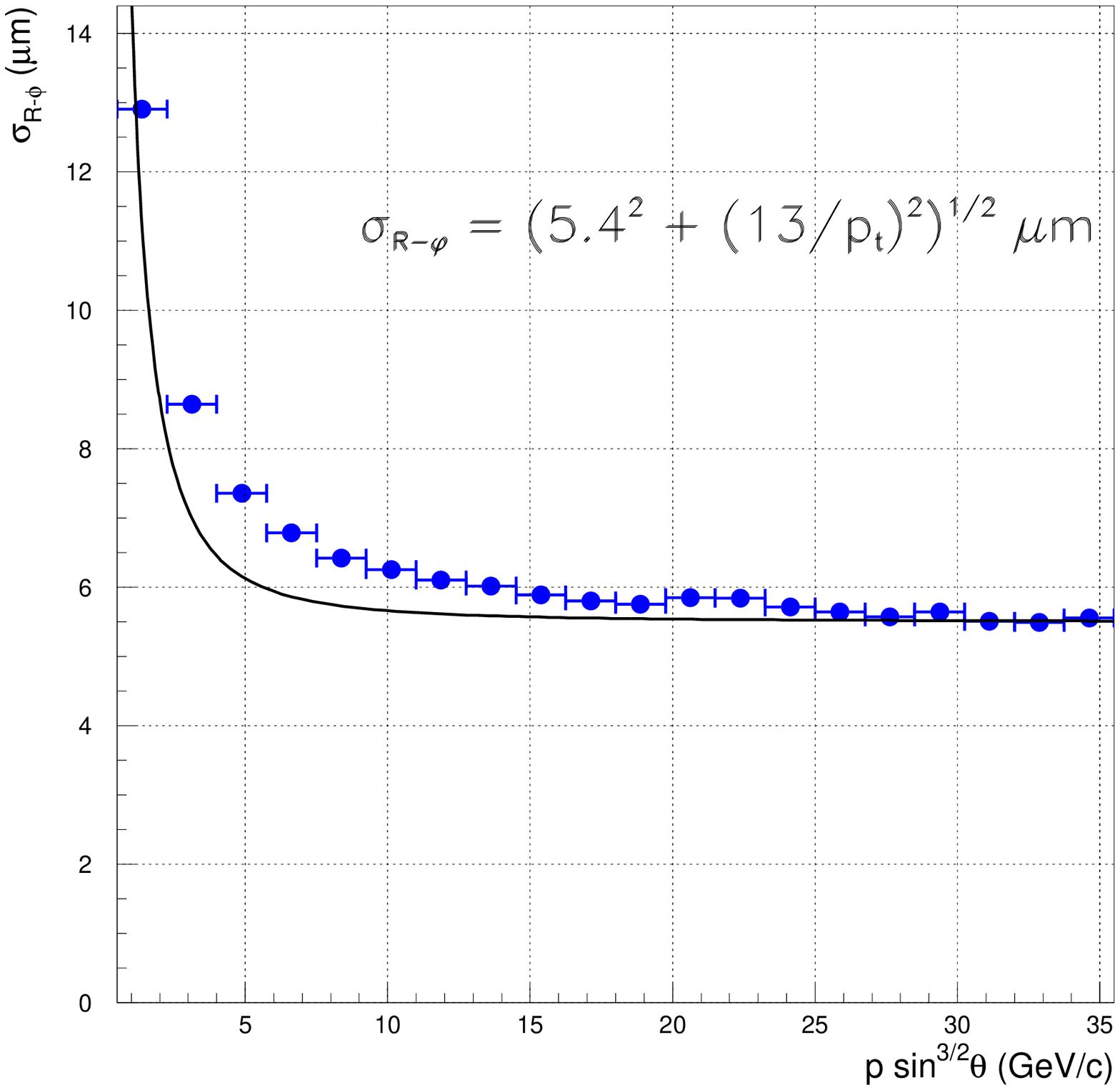,width=7.5cm} &
\epsfig{file=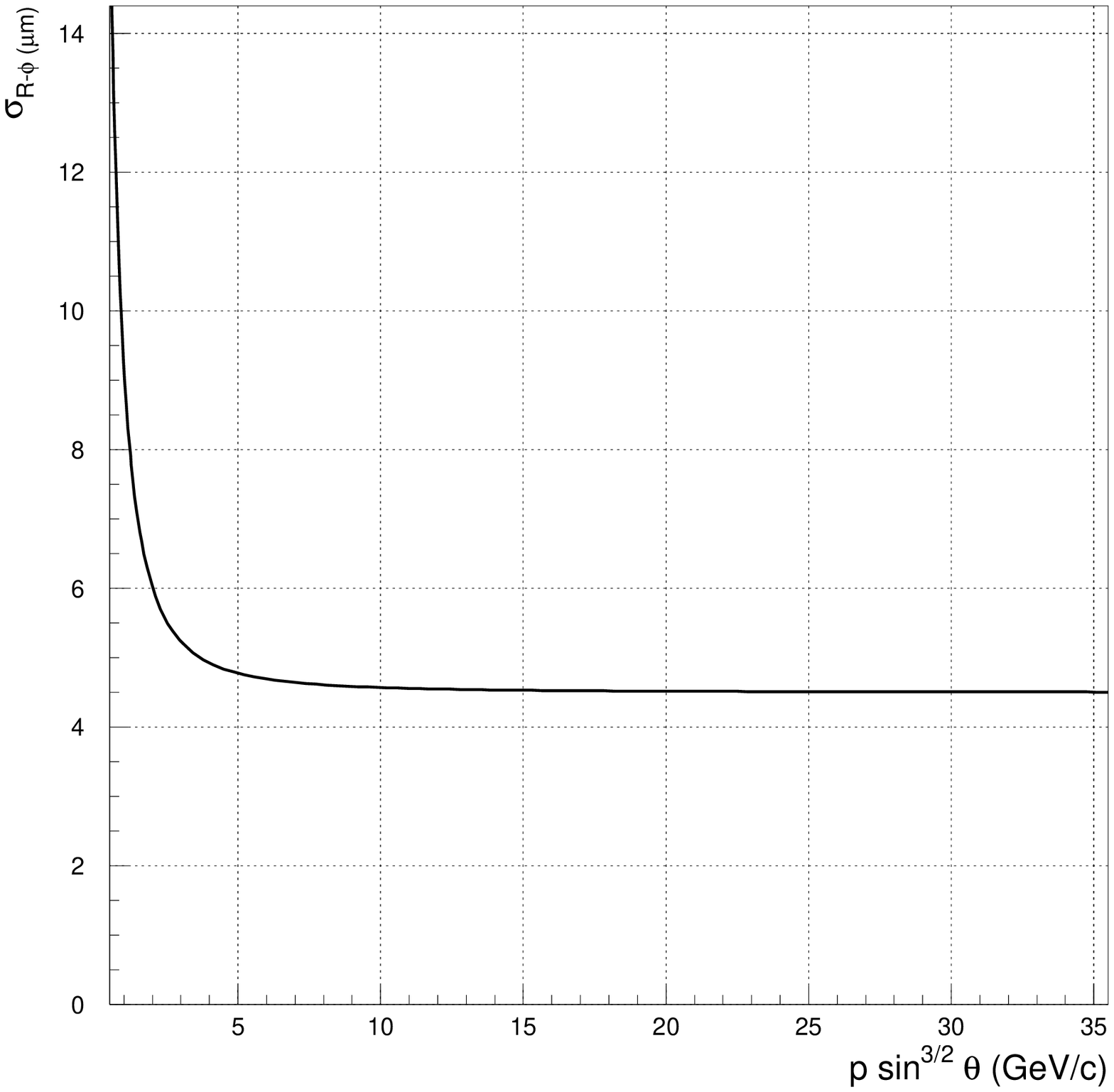,width=7.5cm} \\
\end{tabular}
\end{center}
\captive{The estimated impact parameter resolution for the Hybrid Pixel (left)
and the CCD Vertex Tracker (right), obtained assuming a single point 
resolution of 7.0~$\mu$m and 3.5~$\mu$m respectively.}
\label{fig:ipres}
\end{figure}

Jet flavour tagging is based on the combination of discriminating variables
sensitive to the decay kinematics, the secondary vertex invariant mass and
its topology. The combination can be performed defining a combined likelihood,
by means of Fischer discriminant or by neural networks exploiting non-linear 
correlations between the variables. The typical performance for the most 
demanding flavour tagging task, i.e. the identification of charm, is given in
Figure~1. The jet flavour tagging response can be used either as a cut variable
for event classification, as an input to a global probability defining the 
likelihood of the event to be either signal or background or, finally, as 
fit variable to measure the fraction of decays of a particle, whose properties
are under study, in a given flavour configuration.

The impact of the Vertex Tracker on the physics studies at the linear 
collider can be exemplified by the results on the determination of the 
Higgs decay branching ratios. For its own importance in the study of the 
properties of the Higgs boson and also for its requirements to jet flavour
identification, this process has been adopted as one of the main benchmark
reaction for the optimisation of the Vertex Tracker design. 
The scenarios with a 120~GeV/c$^2$ or a 140~GeV/c$^2$ neutral Higgs boson 
have been studied for an integrated luminosity of 500 fb$^{-1}$ at 
$\sqrt{s}$ = 350~GeV and 500~GeV. 
A jet flavour tagging algorithm has been applied to the di-jets originating 
from the Higgs decay in the
Higgsstrahlung process $e^+e^- \rightarrow Z^0H^0$. The fractions of 
$b \bar b$, $c \bar c$ and $g g$ decays have been extracted by a binned 
maximum likelihood fit to the jet flavour tagging probability distribution.
The results are summarised in Figure~11. The estimated accuracy
allows to distinguish the nature of a neutral Higgs boson, telling
a Minimal Supersymmetric Standard Model (MSSM) Higgs from the Standard Model 
one, for values of the heavy $A^0$ boson mass up to $\simeq$~600~GeV/c$^2$ 
independent on the value of $\tan \beta$~\cite{higgs3,richard}. 
This region is particularly important because it corresponds to the portion of 
the MSSM parameter space where the LHC experiments may only be able to 
observe one Higgs boson without establishing its Supersymmetric or Standard
Model nature. 
\begin{figure}[hb!]
\begin{center}
\epsfig{file=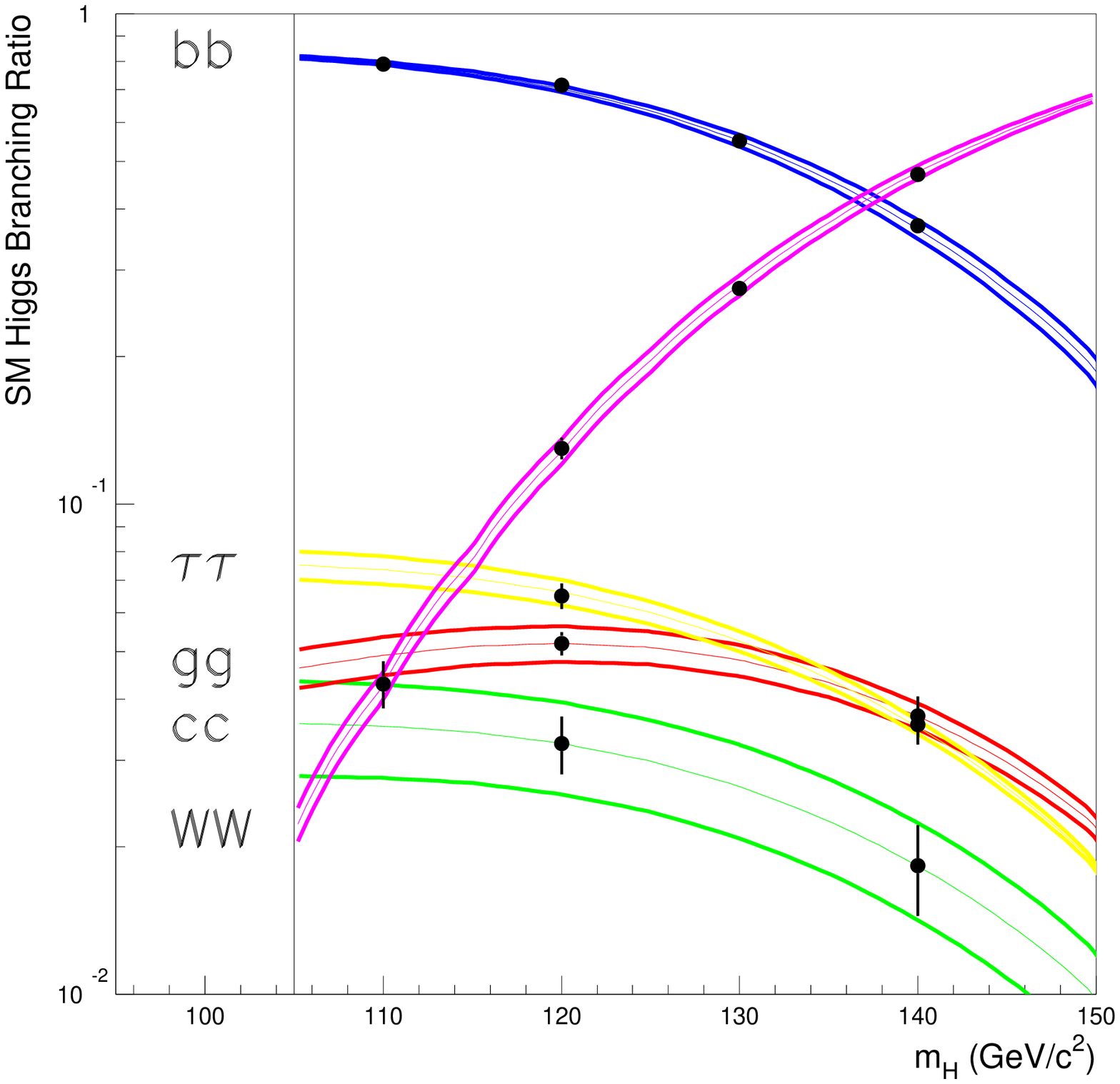,height=9.5cm}
\end{center}
\captive{The predicted SM Higgs decay branching ratios shown as the 68\% 
confidence level bands with overlayed the measured points using the results of
this study (from Ref.~\cite{higgs3}).}
\label{fig:hbr}
\end{figure}

\section{Conclusions}

The linear collider physics programme defines challenging requirements for
the Vertex Tracker. The solutions presently considered for the TESLA project,
rely on the experience with the LEP/SLC detectors, the results of the LHC R\&D
and those of dedicated activities. The estimates for the backgrounds at the 
interaction region and their degree of uncertainty motivate different 
sensor technology to be studied. Pixel sensors guarantee a stable
operation in the interaction region environment however they presently lack the
necessary accuracy and thinness. In order to overcome these limitations R\&D on
hybrid sensors of new design, with interleaved pixels, and monolithic CMOS 
sensors is being carried out. 
CCD devices offer the most attractive performances in this 
respect while they require further R\&D to demonstrate their full 
compatibility with the accelerator timing and backgrounds. The availability of
alternative detector technologies allows to emphasise different aspects of 
experimentation at the linear collider. The tracker design and the detector 
technology will evolve with the results of R\&D, physics studies and updates 
of accelerator parameters towards the definition of the most performant Vertex
Tracker ever designed. Preliminary evaluations of the performance and of the
physics reach define a rich variety of signals of new physics and of
precision measurements relying on the Vertex Tracker resolution.

\section[*]{Acknowledgements}

We thank our collegues of the Helsinki, Milano, 
Krakow and Strasbourg groups for their contributions to the different stages 
of the studies reviewed in this paper. We want to express our gratitude to 
Chris Damerell for many stimulating discussions and his valuable suggestions. 
One of the authors (M.B.) is also thankful to Goran Jarlskog and Leif Jonsson 
for their kind invitation to the Lund Workshop.


\begin{thebibliography}{99}
\bibitem{heuer}
For a review of the general detector concept and its expected 
performances see R.~Heuer, contribution to these Proceedings.
\end{thebibliography}
\end{document}